\documentclass[a4paper,11pt]{article}
\pdfoutput=1 % if your are submitting a pdflatex (i.e. if you have
\usepackage{jcappub} % for details on the use of the package, please
\usepackage[T1]{fontenc} % if needed
\usepackage{bm}
\usepackage{graphicx,subfigure,color}
\usepackage{amssymb,amsmath}
\usepackage{slashed}
\usepackage{array}
\usepackage{hyperref}

\newcommand{\bea}{\begin{eqnarray}}
\newcommand{\eea}{\end{eqnarray}}

\title{\boldmath Feeble DM-SM Interaction via New Scalar and Vector Mediators in Rotating Neutron Stars}

\author[a,1]{Atanu Guha,\note{Corresponding author}}
\author[b]{Debashree Sen}
\affiliation[a]{Department of Physical Sciences\\
Indian Institute of Science Education and Research Pune,\\
Dr. Homi Bhabha Road, Ward No. 8, NCL Colony, Pashan, Pune, India}
\affiliation[b]{Physics Group, Variable Energy Cyclotron Centre, 1/AF Bidhan Nagar, Kolkata 700064, India}
\emailAdd{atanu.guha@students.iiserpune.ac.in}
\emailAdd{d.sen@vecc.gov.in}
\abstract{We investigate the possible presence of dark matter (DM) in massive and rotating neutron stars (NSs). For the purpose we extend our previous work \cite{Sen:2021wev} to introduce a light new physics vector mediator besides a scalar one in order to ensure feeble interaction between fermionic DM and $\beta$ stable hadronic matter in NSs. The chosen masses of DM fermion and the mediators and the couplings are consistent with the self-interaction constraint from Bullet cluster and from present day relic abundance. Assuming that both the scalar and vector mediators contribute equally to the relic abundance, we compute the equation of state (EoS) of the DM admixed NSs to find that the present consideration of the vector new physics mediator do not bring any significant change to the EoS and static NS properties of DM admixed NSs compared to the case where only the scalar mediator was considered \cite{Sen:2021wev}. However, the obtained structural properties in static conditions are in good agreement with the various constraints on them from massive pulsars like PSR J0348+0432 and PSR J0740+6620, the gravitational wave (GW170817) data and the recently obtained results of NICER experiments for PSR J0030+0451 and PSR J0740+6620. We also extend our work to compute the rotational properties of DM admixed NSs rotating at different angular velocities. The present results in this regard suggest that the secondary component of GW190814 may be a rapidly rotating massive DM admixed NS. The constraints on rotational frequency from pulsars like PSR B1937+21 and PSR J1748-2446ad are also satisfied by our present results. Also, the constraints on moment of inertia are satisfied considering slow rotation. The universality relation in terms of normalized moment of inertia also holds good with our DM admixed EoS.}

\begin{document}
\maketitle
\flushbottom

\section{Introduction}
\label{intro}

 Over several years huge endeavor has been made from various perspectives to understand the composition of neutron stars (NSs) and constrain the NS equation of state (EoS) at relevantly high density ($\approx 5-10$ times the normal matter density $\rho ~(\approx 0.16~\rm{fm^{-3}}$)). At such density, the lack of proper understanding of the properties of matter and interactions from experimental perspectives still makes the composition of NS core inconclusive and the NS EoS loosely constrained. However, theoretical research in this direction has acquired a new impetus with the detection of gravitational wave GW170817 \cite{TheLIGOScientific:2017qsa,Abbott:2018exr} from binary NS merger (BNSM) which provided strong constraints on the radius and dimensionless tidal deformability of 1.4 $M_{\odot}$ NS \cite{TheLIGOScientific:2017qsa,Fattoyev:2017jql,De:2018uhw,Most:2018hfd}. In 2019 the detection of GW190425 from another BNSM also put constraints on the mass-radius dependence of NSs \cite{Abbott:2020aai}. In the same year the detection of GW190814 from compact object mergers has still remained baffling because of the mass of the secondary compact object associated with it \cite{Abbott:2020khf}. In absence of any further information related to GW190814 like its electromagnetic counterpart or the tidal deformability of this compact object, it is totally undermined whether this object is a black hole (BH) or a NS. Going with first possibility it will then be the lightest BH detected so far while the second possibility suggests that it may be the most massive NS detected till date. Recent works \cite{Zhang:2020zsc,Biswas:2020xna} suggest that it may also be the fastest rotating pulsar discovered as yet. Both the works have tried to determine and prescribe the possible rotational frequency of the object assuming it to be a NS. On the other hand, by invoking first order hadron-quark phase transition, \cite{Rather:2021yxo} showed that it may be a fast rotating hybrid star while \cite{Tan:2021ahl} proposed that with suitable choice of hadronic EoS, certain structure in the speed of sound can imply that this object could be a non-rotating hybrid star. Apart from the constraints obtained from gravitational wave data analysis, the recent experiments carried out by NICER have come up with some strong constraints on the mass-radius plane of NSs. NICER have not only prescribed the mass and radius values of PSR J0030+0451 \cite{Riley:2019yda,Miller:2019cac} with high accuracy but also estimated the radius of PSR J0740+6620 \cite{Miller:2021qha,Riley:2021pdl} which is the most massive pulsar discovered till date \cite{Fonseca:2021wxt}. Apart from the constraints on NS mass and radius, a few other observational and theoretical constraints on the other properties of NS also help us to constrain the NS EoS to some extent. For example \cite{Sanwal:2002jr} and \cite{Hambaryan:2017wvm} prescribed the surface redshift of 1E 1207.4-5209 and RX J0720.4-3125, respectively from source spectrum analysis. The rotational frequency of rapidly rotating pulsars like PSR B1937+21 \cite{Backer} and PSR J1748-2446ad \cite{Hessels:2006ze} were also determined very accurately. The joint results of PSR J0030+0451 and GW170817 observation and the nuclear data analysis helped to predict the moment of inertia and gravitational binding energy of NS of canonical mass 1.4 $M_{\odot}$ \cite{Jiang:2019rcw}. Also \cite{Landry:2018jyg} found the moment of inertia of PSR J0737-3039A of mass 1.338 $M_{\odot}$. Moreover, in the slow rotational conditions ($P\leq 10s$) the universal relations in terms of normalized moment of inertia, quadrupole, tidal deformability and compactness also set constraints on the NS EoS \cite{Yagi:2013awa,Breu:2016ufb}.

 Therefore it becomes quite challenging to satisfy the aforesaid constraints on NS EoS which is largely determined by the composition and interactions chosen. Literature provides a huge amount of work that predicts the NS EoS considering different forms of matter at such density \cite{Glendenning}. Several works also predicted that besides hadronic matter, different forms of dark matter (DM) may also exist inside the core of NSs \cite{Panotopoulos:2017idn, Bertoni:2013bsa, Nelson:2018xtr,Ellis:2018bkr, Li:2012ii, Tolos:2015qra, Deliyergiyev:2019vti, Rezaei:2016zje, Mukhopadhyay:2016dsg}. Such stars are often referred to as DM admixed NSs. 

  The existence of DM in our known Universe is supported by the observational evidences like the rotation curves of the galaxies, observation of gravitational lensing, X-ray analysis of Bullet cluster~\cite{Bertone:2004pz, Aghanim:2018eyx}. The Cosmic Microwave Background (CMB) anisotropy maps, obtained from the Wilkinson Microwave Anisotropy Probe (WMAP) data \cite{Ade:2013zuv, Bennett:2012zja}, furnishes the present day thermal relic abundances of DM to be $\sim \Omega h^2 \approx 0.12$~ \cite{Tanabashi:2018oca,Plehn:2017fdg,Cannoni:2015wba}. The exact nature and properties of DM particle candidates are not yet known. In literature, several popular models can explain the phenomena that were believed to be caused by DM from cosmological evidences. One of the key feature of a feasible DM model is that the model has to reproduce the observed non-baryonic relic density to be recognized as a successful model. This requirement sets strong constraints on the model parameters. Weakly interacting massive particles (WIMPs) are the most suitable DM particle candidates and the WIMPs were in thermal equilibrium with the plasma in early Universe before decoupling.  
 
 Two major possible ways to explain the possible existence of DM in NSs are that the trapped DM during the supernova (SN) explosion may be inherited by NSs or the later may also accrete DM in its later stages \cite{Razeira:2011zza,Guver:2012ba,PerezGarcia:2010ap,deLavallaz:2010wp,Ciarcelluti:2010ji}. We consider the second possibility in the present work. Accretion or the DM capture by NSs is caused by the scattering of the adjacent DM particles into the NS matter (NSM). DM particles scatter off nucleons/electrons, lose kinetic energy in the process and become gravitationally bound to the NSM~\cite{Bell:2019pyc}. The possibility of multiple scattering before the capture ensures a wide range of mass of DM particle to be potentially available for the accretion~\cite{Bramante:2017xlb,Dasgupta:2019juq,Garani:2020wge}. After getting captured, the DM particles encounter further collisions before they completely lose their kinetic energy and attain thermal equilibrium with the system~\cite{Bell:2019pyc, Bell:2020lmm}. At this stage the DM particles become confined within the radius of the NS~\cite{Garani:2020wge, Catena:2016sfr}. The accretion stops at a point when DM particles get saturated and they are in thermal equilibrium among themselves due to the self interactions, which also ensures nearly a constant density of DM particles within the confined region. From the present understanding, as the DM self-scattering crosssection is much stronger than the DM-nucleon/DM-electron crosssection~\cite{Tulin:2013teo,Dasgupta:2020dik,Kouvaris:2014uoa}, the equilibrium density of the captured DM particles is very less compared to the saturated matter density in order to maintain the equilibrium among themselves as well as with the nucleons. Several works have successfully explained the possible existence of DM admixed NSs in which the hadronic matter and DM do not interact \cite{Ellis:2018bkr, Li:2012ii, Tolos:2015qra, Deliyergiyev:2019vti, Rezaei:2016zje, Mukhopadhyay:2016dsg} while the interaction of the DM particles with the baryons via exchange of Higgs boson has also been considered by \cite{Panotopoulos:2017idn, Bertoni:2013bsa, Nelson:2018xtr, Bhat:2019tnz,%Das:2020vng,
Das:2018frc,Quddus:2019ghy}%,Das:2020ptd}
for the same purpose. We invoke feeble interaction between $\beta$ equilibrated hadronic NSM and fermionic DM. For the purpose we proposed the exchange of a new physics scalar mediator that mixes with SM Higgs boson \cite{Sen:2021wev}. It is well known that in the hadronic sector the vector mesons are responsible for generating the repulsive effects which contribute considerably to the EoS \cite{Glendenning, Boguta, Sahu:2000ut,Jha:2009kt}. Motivated by this fact, in the present work we include the vector new physics mediator in addition to the scalar one to study the net impact of the feeble DM-SM interaction on the DM admixed EoS and consequently the structural properties of the DM admixed NS in both static and rotating conditions. 
 
 The structural properties of NSs are largely dependent on the EoS which in turn is determined by the composition and interaction of different particles considered. The extreme conditions of matter characterized by high compactness and low temperature at supra-nuclear densities, relevant to NS cores, makes it highly challenging to attain from experimental perspectives. Thus the composition of NS core is largely inconclusive from experimental perspectives and one therefore relies on the theoretical modeling of NSM to understand the composition, dynamics and structure of NSs. So the theoretical calculations of the structural properties of NSs are largely model dependent but the validity of such results can be tested in accordance with the various observational constraints on the different structural properties of NSs (as discussed in the first paragraph of this section). Literature suggests that a huge number of theoretical models, considering different possible composition of NSs, have been applied to obtain the EoS and the structural properties of NSs. Of them \cite{Ellis:2018bkr, Li:2012ii, Tolos:2015qra, Deliyergiyev:2019vti, Rezaei:2016zje, Mukhopadhyay:2016dsg, Panotopoulos:2017idn, Bertoni:2013bsa, Nelson:2018xtr} have investigated the possible presence of DM in NSs considering particular hadronic models. Similar to such works, we adopt the effective chiral model \citep{Sahu:2000ut,Jha:2009kt} to account for the hadronic part of the DM admixed NSM.
 
 The present paper is organized as follows. In section \ref{formalism} we present the basic details of the model in presence of DM and the parameter sets considered for both hadronic matter \ref{Param_HM} and DM \ref{Param_DM}. Numerical results and corresponding discussions can be found in section \ref{Results}. Finally we conclude
in section \ref{Conclusion}.

\section{Formalism}
\label{formalism}

\subsection{Model including scalar and vector new physics mediator}

 In our previous work~\cite{Sen:2021wev} we invoked feebly interacting dark sector through a higgs portal scalar new physics mediator, which interacts with both SM and DM fermions in NSM. For the hadronic matter sector we considered the effective chiral model \cite{Sahu:2000ut,Jha:2009kt} which is based on chiral symmetry with the pseudoscalar $\pi$ and scalar $\sigma$ as chiral partners and $x^2=(\pi^2 + \sigma^2)$. In the pure hadronic sector the nucleons $\psi$ interact via the scalar $\sigma$, vector $\omega$, isovector $\rho$ mesons as mediators with respective couplings $g_{\sigma}$, $g_{\omega}$ and $g_{\rho}$. The pions do not contribute in the mean field approximation \cite{Sahu:2000ut,Jha:2009kt,Sen:2018yyq,Sen:2020edi,Sen:2021bms}. In addition to hadronic matter, for the dark sector, we consider a phenomenological approach to describe the self-interaction of non-relativistic DM by a Yukawa potential~\cite{Tulin:2013teo}. 
 
\bea
V(r) = \pm \frac{\alpha_\chi}{r} e^{-m_\phi r}
\eea

where, $\alpha_\chi = \frac{y^2}{4 \pi}$ is the dark fine structure constant.

 This interaction can be both attractive and repulsive. In the present work, we assume that in the dark sector, in addition to the scalar mediator $\phi$, a vector mediator $\xi$ to be also present in the theoretical framework (like the $\sigma$ and $\omega$ mesons in the hadronic sector) having respective couplings as $y_{\phi}$ and $y_{\xi}$ with the fermionic DM.

\bea
\mathcal{L}_{int} = \begin{cases}
y_\phi \phi \bar{\chi} \chi \\
y_\xi \bar{\chi} \gamma_\mu \chi \xi^\mu
\end{cases} 
\eea

 Scalar interactions are purely attractive while vector interactions are both attractive ($\chi \bar{\chi}$ scattering) and repulsive ($\chi \chi$ and $\bar{\chi} \bar{\chi}$ scattering).
 
% Let us consider the dark sector contains a broken $U(1)'$ symmetry and the new $U(1)'$ gauge boson is kinetically mixed SM $\omega$ meson~\cite{Mahoney:2017jqk}.
% 
%\bea
%\mathcal{L}_{vec} = y'_\xi \bar{\chi} \gamma^\mu \tilde{\xi}_\mu \chi + g'_\xi \bar{\psi} \gamma^\mu \tilde{\omega}_\mu \psi
%\eea
%
%where, $\tilde{\xi}$ is the dark gauge boson and $\tilde{\omega}$ is the SM gauge boson. Both are the admixture of $\xi$ and $\omega$ meson.
%
%Due to the kinetic mixing term $\frac{1}{2} \frac{\epsilon}{\sqrt{1+ \epsilon^2}} \tilde{F}_{\mu \nu} \tilde{V}^{\mu \nu}$, $\xi$ can in principle obtain effective coupling to the nucleons. The viable parameter range is $\epsilon \ll 1$. Therefore,
%
%\bea
%m_\xi = m_{\tilde{\xi}} \sqrt{1+\epsilon^2} \simeq m_{\tilde{\xi}} \nonumber \\
%m_\omega = m_{\tilde{\omega}} \sqrt{1+\epsilon^2} \simeq m_{\tilde{\omega}}
%\eea 

The new scalar mediator obtain an effective coupling $g_{\phi NN}$ with the nucleon through the higgs portal mixing~\cite{Sen:2021wev}. In the present discussion, we assume that the new vector mediator also obtain an effective coupling $g_{\xi}$ to the nucleons through the kinetic portal mixing with the SM mesons~\cite{Mahoney:2017jqk}. With the assumption that $g_{\xi}$ is of the order of $g_{\phi NN}$, we can write the total Lagrangian density as
 
 \bea
 \mathcal{L} &=& \bar{\psi} \left[ \left( i \gamma_\mu \partial^\mu - g_\omega \gamma_\mu \omega^\mu - \frac{1}{2} g_\rho \vec{\rho_\mu}\cdot \vec{\tau} \gamma^\mu - g_\xi \gamma_\mu \xi^\mu \right) - g_\sigma \left( \sigma + i \gamma_5 \vec{\tau}\cdot \vec{\pi} \right) -g_{\phi N N} \phi \right] \psi \nonumber \\
 &+& \frac{1}{2} \left(\partial_\mu \vec{\pi} \partial^\mu \vec{\pi} + \partial_\mu \sigma \partial^\mu \sigma \right) - \frac{\lambda}{4} \left(x^2 - x_0^2 \right)^2 - \frac{\lambda B}{6} \left(x^2 - x_0^2 \right)^3 - \frac{\lambda C}{8} \left(x^2 - x_0^2 \right)^4 \nonumber \\
 &-& \frac{1}{4} F_{\mu \nu} F^{\mu \nu} + \frac{1}{2} g_\omega^2 x^2 \omega_\mu \omega^\mu - \frac{1}{4} \vec{R}_{\mu \nu} \cdot \vec{R}^{\mu \nu} + \frac{1}{2} m_\rho^2 \vec{\rho_\mu}\cdot \vec{\rho^\mu} -\frac{1}{4} V_{\mu \nu} V^{\mu \nu} +\frac{1}{2} m_\xi^2 \xi_\mu \xi^\mu  \nonumber \\
 &+& \frac{1}{2} \partial_\mu \phi \partial^\mu \phi - \frac{1}{2}m_\phi^2 \phi^2  +  \bar{\chi} \left[\left( i \gamma_\mu \partial^\mu - y_\xi \gamma_\mu \xi^\mu \right) - \left(m_\chi + y_\phi \phi \right) \right] \chi 
 \label{Eq:Lagrangian}
 \eea
 
 $\psi$, being the nucleon isospin doublet, $\overline{\psi}=\psi^{\dagger} \gamma_0$. $\vec{\tau}$ and $\gamma^{\mu}$ are the Pauli and Dirac matrices, respectively. The kinetic terms for the $\omega$ and $\rho$ fields are $\frac{1}{4} F_{\mu \nu} F^{\mu \nu}$ and $\frac{1}{4} \vec{R}_{\mu \nu} \cdot \vec{R}^{\mu \nu}$, respectively, where $F_{\mu\nu}=\partial_\mu \omega_\nu - \partial_\nu \omega_\mu$ and $\vec{R}_{\mu\nu}=\partial_\mu \vec{\rho}_\nu - \partial_\nu \vec{\rho}_\mu$. Here $B$ and $C$ are coefficients of the higher order scalar field terms. Similarly, $\frac{1}{4} V_{\mu \nu} V^{\mu \nu}$ is the kinetic term associated with the vector mediator $\xi$ from the dark sector with $V_{\mu\nu}=\partial_\mu \xi_\nu - \partial_\nu \xi_\mu$. Theoretically, at high density relevant to NS cores, there may be possibility of existence of exotic forms of matter like the hyperons and delta baryons \cite{Glendenning,Weissenborn:2011kb,Miyatsu:2011bc,vanDalen:2014mqa,Drago:2014oja,Li:2019tjx,Sen:2021bms}, boson condensates \cite{Kolomeitsev:1995xz,Ramos:2000dq,Malik:2021nas} and even deconfined quark matter \cite{Glendenning,Weissenborn:2011qu,Ozel:2010bz,Klahn:2013kga,Drago:2015cea,Wu:2017xaz,Alford:2019oge,Sen:2021bms,Sen:2021cgl}. Such exotics are known to soften the equation of state and reduce the maximum mass of the neutron stars. However, in absence of any concrete experimental or observational evidence to support the presence of such exotic forms of matter in NSs, we have considered the most fundamental and widely considered $\beta$ equilibrated matter as the hadronic matter composition \cite{Bhat:2019tnz,%Das:2020vng,
Das:2018frc,Quddus:2019ghy,%Das:2020ptd,
Li:2012qf,Ellis:2018bkr,Tolos:2015qra, Deliyergiyev:2019vti, Rezaei:2016zje, Mukhopadhyay:2016dsg}. The terms $\bar{\psi}(g_{\phi N N} \phi)\psi$ and $\bar{\psi}(g_\xi \gamma_\mu \xi^\mu)\psi$ of the above equation \ref{Eq:Lagrangian} indicate the interaction of the the nucleons $\psi$ with the scalar $\phi$ and the vector $\xi$ new physics mediators from dark sector. Like \cite{Panotopoulos:2017idn} we have taken attractive potential for the new physics scalar mediators consistent with \cite{Hambye:2019tjt,Barbieri:1988ct}. The interaction between fermionic DM $\chi$ with the DM mediators $\phi$ and $\xi$ are depicted by the last term of equation \ref{Eq:Lagrangian}.

 Due to chiral symmetry breaking, the mass of nucleons $m$ and that of the scalar $m_{\sigma}$ and the vector $m_{\omega}$ mesons are obtained in terms of the vacuum expectation value (VEV) of the scalar field $\left\langle \sigma \right\rangle = x_0$ as

\bea
m = g_{\sigma} x_0 + g_{\phi NN} \phi_0; ~~~ m_{\sigma} = \sqrt{2 \lambda}~ x_0; ~~~ m_{\omega}=g_{\omega} x_0 
\eea

 Note that due to interaction with the DM sector via scalar $\phi$ mediator, the nucleon mass $m$ is now a function of VEVs of both the scalar fields $x_0$ and $\phi_0$.
 
 The scaled vector and scalar meson couplings of the pure hadronic sector are 
 
\bea
C_{\omega} = \frac{g^2_{\omega}}{m^2_{\omega}};~~~ C_{\sigma} = \frac{g^2_{\sigma}}{m^2_{\sigma}}
\eea

 The effective masses of the nucleons ($m^\star$) and DM ($m^\star_\chi$) are obtained as
 
\bea
m^\star &=& g_\sigma \sigma + g_{\phi N N} \phi \nonumber \\
m^\star_\chi &=& m_\chi + y_\phi \phi
\eea

 The equation of motion for the nucleons and different mesons of both the hadronic and dark sectors can be obtained by applying the mean field treatment.
 
  The scalar equation of motion is now $\phi_0$ dependent and in terms of $Y=m^{\star}/m$ it is given as
 
\bea
(1 - Y^2) -\frac{B}{C_{\omega}}(1-Y^2)^2 +\frac{C}{C_{\omega}^2}(1-Y^2)^3 -\frac{2C_{\sigma}\rho_s}{Y(m - g_{\phi NN}\phi_0)} +\frac{2C_{\sigma}C_{\omega}}{Y^4(m - g_{\phi NN}\phi_0)^2} = 0 \nonumber \\
\label{scalar_field}
\eea

while the vector $\omega$ and isovector $\rho$ meson field equations remain unchanged and same as that obtained without including DM \cite{Sahu:2000ut,Jha:2009kt} viz.

\bea
\omega_0=\frac{\rho}{g_{\omega} x^2}
\label{vector_field}
\eea

and
\bea
\rho_{03}=\sum_{N}\frac{g_{\rho}}{m_\rho^2}I_{3_N}\rho_N 
\label{isovector_field}
\eea

Here `3' denotes for the third component in isospin $I$ of the individual nucleons $N$. 

 The scalar density is given as
 
\bea
\rho_S=\left\langle\overline{\psi}\psi\right\rangle=\frac{\gamma}{2 \pi^2} \int^{k_F}_0 dk ~k^2 \frac{m^*}{\sqrt{k^2 + {m^{*}}^2}}
\eea

 with $k_F$ be the Fermi-momentum; while the baryon density as

\bea
\rho=\left\langle\psi^\dagger\psi\right\rangle=\frac{\gamma}{2\pi^2} \int^{k_F}_0 dk ~k^2  
\eea

 Here the total baryon density is the sum of individual nucleon densities i.e., $\rho=\rho_n + \rho_p$ and $\gamma$ denotes the spin degeneracy factor. For symmetric nuclear matter (N=Z) $\gamma=$ 4 while for asymmetric nuclear matter $\gamma=$ 2.

 The equation of motion of the new scalar mediator field is given as

\bea
\phi_0=\frac{m^\star_\chi-m_\chi}{y_{\phi}}
\eea

while that of the new vector mediator field is

\bea
\xi_0 = \frac{g_\xi \rho + y_\xi \rho_\chi}{m_\xi^2}
\eea

 where, $\rho_\chi$ is the density of the DM fermions and $m_\xi$ is the mass of the new vector mediator.

\subsection{Equation of State}
 
 The EoS viz. the energy density $\varepsilon$ and pressure $P$ is obtained by calculating the energy-momentum tensor involving the Lagrangian density \ref{Eq:Lagrangian}. The total energy density is computed as

\bea
\varepsilon &=& \frac{\left(m - g_{\phi NN}\phi_0\right)^2 \left(1-Y^2\right)^2}{8 C_{\sigma}} - \frac{\left(m - g_{\phi NN}\phi_0\right)^2 B \left(1-Y^2\right)^3}{12 C_{\sigma} C_{\omega}} \nonumber \\
&+& \frac{\left(m - g_{\phi NN}\phi_0\right)^2 C \left(1-Y^2\right)^4}{16 C_{\sigma} C_{\omega}^2} + \frac{C_{\omega} \rho^2}{2 Y^2}  \nonumber \\
&+&  \frac{1}{2} m_\rho^2 \rho_{03}^2 + \frac{\gamma}{2 \pi^2} \int_0^{k_F} \sqrt{k^2 + {m^{\star}}^2}~ k^2 dk + \frac{\gamma}{2 \pi^2} \sum_{\lambda^\prime=e,\mu} \int_0^{k_{\lambda^\prime}}  \sqrt{k^2 + m_{\lambda\prime}^2}~ k^2 dk  \nonumber \\
&+& \frac{1}{2} m_\phi^2 \phi_0^2 + \frac{1}{2} m_\xi^2 \xi_0^2 + \frac{\gamma_\chi}{2 \pi^2} \int_0^{k_F^{\chi}} \sqrt{k_\chi^2 + {m^{\star}}_\chi^2}~ k_\chi^2 dk_\chi
\label{e}
\eea 

and the total pressure as

\bea
P &=& - \frac{\left(m - g_{\phi NN}\phi_0\right)^2 \left(1-Y^2\right)^2}{8 C_{\sigma}} + \frac{\left(m - g_{\phi NN}\phi_0\right)^2 B \left(1-Y^2\right)^3}{12 C_{\sigma} C_{\omega}} \nonumber \\ 
&-& \frac{\left(m - g_{\phi NN}\phi_0\right)^2 C \left(1-Y^2\right)^4}{16  C_{\sigma} C_{\omega}^2} + \frac{C_{\omega} \rho^2}{2 Y^2} \nonumber \\
&+&  \frac{1}{2} m_\rho^2 \rho_{03}^2 + \frac{\gamma}{6 \pi^2} \int_0^{k_F} \frac{k^4 dk}{\sqrt{k^2 + {m^{\star}}^2}} + \frac{\gamma}{6 \pi^2} \sum_{\lambda^\prime=e,\mu} \int_0^{k_{\lambda^\prime}}  \frac{k^4 dk}{\sqrt{k^2 + m_{\lambda^\prime}^2}}  \nonumber \\
&-& \frac{1}{2} m_\phi^2 \phi_0^2 + \frac{1}{2} m_\xi^2 \xi_0^2 + \frac{\gamma_\chi}{6 \pi^2} \int_0^{k_F^{\chi}} \frac{k_\chi^4 dk_\chi}{\sqrt{k_\chi^2 + {m^{\star}}_\chi^2}}
\label{P}
\eea 
 
\subsection{Model parameter set for hadronic matter}
\label{Param_HM}

The five parameters of the hadronic model ($C_i=g_i^2/m_i^2$ where $i=\sigma,\omega,\rho$ and $B$ and $C$) are determined by reproducing the saturated nuclear matter (SNM) properties. The hadronic model parameter set, chosen for the present work, is adopted from \cite{Jha:2009kt} and is given in table \ref{table_NM} along with the SNM properties corresponding to the chosen set.

\begin{table}[!ht]
\caption{Parameter set for the hadronic model (adopted from \cite{Jha:2009kt}) along the saturation properties.}
{{
\setlength{\tabcolsep}{3.5pt}
\centering
\begin{tabular}{cccccccccccccc}
\hline
\hline
%\multicolumn{1}{c}{Model}&
\multicolumn{1}{c}{$C_{\sigma}$}&
\multicolumn{1}{c}{$C_{\omega}$} &
\multicolumn{1}{c}{$C_{\rho}$} &
\multicolumn{1}{c}{$B/m^2$} &
\multicolumn{1}{c}{$C/m^4$} &
\multicolumn{1}{c}{$Y$} &
\multicolumn{1}{c}{$m_{\sigma}$} &
\multicolumn{1}{c}{$f_{\pi}$} &
\multicolumn{1}{c}{$K$} & 
\multicolumn{1}{c}{$B/A$} &
\multicolumn{1}{c}{$J(L_0)$} &
\multicolumn{1}{c}{$\rho_0$} \\
%
%\multicolumn{1}{c}{ } &
\multicolumn{1}{c}{($fm^2$)} &
\multicolumn{1}{c}{($fm^2$)} &
\multicolumn{1}{c}{($fm^2$)} &
\multicolumn{1}{c}{($fm^2$)} &
\multicolumn{1}{c}{($fm^4$)}&
\multicolumn{1}{c}{} &
\multicolumn{1}{c}{(MeV)} &
\multicolumn{1}{c}{(MeV)} &
\multicolumn{1}{c}{(MeV)} &
\multicolumn{1}{c}{(MeV)} &
\multicolumn{1}{c}{(MeV)} &
\multicolumn{1}{c}{($fm^{-3}$)} \\
\hline
\hline

7.325  &1.642  &5.324 &-6.586   &0.571    &0.87  &444.614   &153.984  &231  &-16.3   &32(88)  &0.153 \\ 
\hline
\hline
\end{tabular}
}}
\protect\label{table_NM}
\end{table}

 The values of SNM properties like the symmetry energy coefficient ($J = 32$~MeV), saturation density ($\rho_0 = 0.153$~$\rm{fm^{-3}}$), binding energy per particle ($B/A = -16.3$~MeV) and the nuclear incompressibility ($K = 231$~ MeV), yielded by the model parameter, agree well with constraints from \cite{Stone:2006fn,Dutra:2014qga,Khan:2012ps,Khan:2013mga,Garg:2018uam}. The value slope parameter ($L_0 = 87$~MeV) lie within the range prescribed by \cite{Dutra:2014qga} and is also comparable with the results of \cite{Fattoyev:2017jql,Zhu:2018ona} based on the co-relations between the symmetry energy and tidal deformability and radius of a 1.4 $M_{\odot}$ NS. The same parameter set was also used in our previous work \cite{Sen:2021wev} to invoke feeble SM-DM interaction via dark scalar mediator in NSM.

 We next present the parameter sets chosen for the dark sector in section \ref{Param_DM}.

\subsection{Parameter set for dark sector}
\label{Param_DM}

 The structural analysis of the bullet cluster is one of the most subtle evidence of the existence of DM. It not only supports the abundance of DM at the cluster scale, but also provides an estimate of the self-interaction between the DM particle candidates~\cite{Randall:2007ph, Bradac:2006er}. We consider some of the combinations of the mass ($m_\chi$) of the DM fermions ($\chi, \bar{\chi}$) and the corresponding values of the masses ($m_\phi, m_\xi$) of the light mediators $\phi$ and $\xi$ as benchmark points which satisfy the self-interaction constraints from bullet cluster~\cite{Tulin:2013teo, Tulin:2017ara,Hambye:2019tjt}. With the knowledge of the present day thermal relic abundances of DM, we determine the values of the couplings $y_\phi$ and $y_\xi$~\cite{Belanger:2013oya, Gondolo:1990dk, Guha:2018mli}. The parameter sets for the dark sector have been shown in Table~\ref{table_DM}. In this work we consider the production of non-thermal DM fermions ($\chi, \bar{\chi}$) inside the NS to be negligible.

\begin{table}[ht!]
\caption{Chosen values of self interacting DM $m_\chi$  and corresponding values of $m_\phi$ and $m_\xi$ from the constraints obtained from Bullet cluster. $y_\phi$ and $y_\xi$ have been fixed from observed relic abundance.}
{{
\setlength{\tabcolsep}{25.5pt}
\begin{center}
\begin{tabular}{ c c c c c c c c}
\hline
\hline
 $m_{\chi}$ & $m_{\phi}$ & $m_{\xi}$ & $y_{\phi}$ & $y_{\xi}$ \\
 (GeV) &(MeV) &(MeV)  &  \\

\hline
\hline
%$1$ & $4$ & $5$ & $0.07$ & $0.07$ \\
$5$ & $9$ & $11$ & $0.13$ & $0.13$ \\
%$10$ & $15$ & $19$ & $0.18$ & $0.18$ \\  
$15$ & $20$ & $34$ & $0.21$ & $0.21$ \\  
%$20$ & $30$ & $53$ & $0.24$ & $0.24$ \\  
%$50$ & $60$ & $77$ & $0.38$ & $0.38$ \\  
\hline\hline
\end{tabular}
\end{center}
}}
\protect\label{table_DM}
\end{table}
 
\begin{figure}[!ht]
\centering
\includegraphics[width=0.5\textwidth]{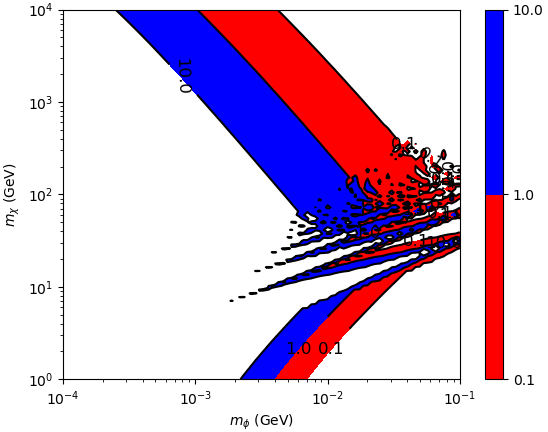}\hfill
\includegraphics[width=0.5\textwidth]{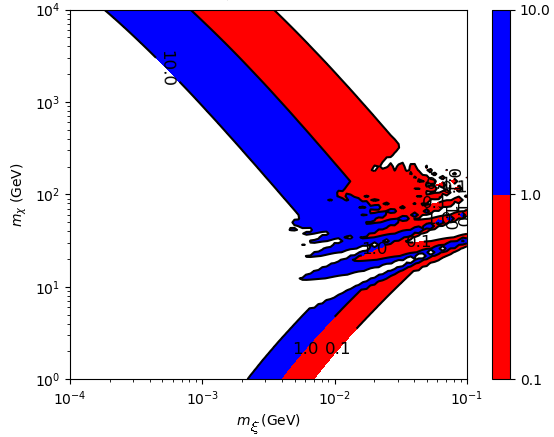}
\caption{Combination of $m_\chi,m_\xi$ and $m_\phi$ satisfying the self-interaction constraint from bullet cluster~\protect\cite{Randall:2007ph, Tulin:2013teo}. The color coding denotes the values of DM self-interaction $\sigma_T/m_\chi = (0.1\rm{-}1)~\rm{cm^2/gm}$ (red) and $(1\rm{-}10)~\rm{cm^2/gm}$ (blue).}
\label{bullet_cluster}
\end{figure}

 The independent couplings $y_\phi$ and $y_\xi$ have been fixed from the observed relic abundance~\cite{Belanger:2013oya} considering that both the channels via the scalar and the vector mediators, contribute equally to the relic density. In figure \ref{bullet_cluster} we showed the region $\sigma_T/m_\chi = (0.1\rm{-}10)~\rm{cm^2/gm}$ ($\sigma_T$ be the self-scattering transfer cross-section) at both the panels for scalar and vector mediators. This is the typical range of values of $\sigma_T$ for the known galaxies and clusters~\cite{Randall:2007ph,Bradac:2006er,Dawson:2011kf,Dave:2000ar,Vogelsberger:2012ku,Kahlhoefer:2015vua}. The constraints from bullet cluster structure formation states that $\sigma_T/m_\chi \leq 1.25~\rm{cm^2/gm}$~\cite{Randall:2007ph, Robertson:2016xjh} for DM fermions of mass $m_\chi$, self-scattered through the light scalar(vector) mediators of mass $m_\phi$($m_\xi$).   
%%%%%%%%%%%%%%%%%%%%%%%%%%%%%%  
%%%%%%%%%%%%%%%%%%%%%%%%%%%%%%%%%%%% 

 Interaction between the dark sector and the nucleons is mediated by the new physics mediators. The corresponding mediator-nucleon couplings are much smaller compared to the SM interactions which ensures that the order of magnitude of DM-nucleon crosssection does not fall in the excluded region of the parameter space. This is also supported by the WIMP miracle. For the scalar mediator case the coupling has been calculated following~\cite{Sen:2021wev,Cheng:2012qr,Backovic:2015cra} and for the vector mediator case it has been assumed to be the same. For both the cases mediator-nucleon couplings are smaller than the SM Higgs-nucleon coupling.
  
 Certain works \cite{Garani:2020wge,Bell:2019pyc,Bramante:2017xlb,Dasgupta:2019juq} have considered that once the DM particles get captured, they undergo further collisions, lose kinetic energy and their orbits shrink down to the thermal radius, a quantity which is determined by the NS temperature and the core density. Clearly, at equilibrium the kinetic energy of a DM particle equals to the temperature of NS. The exact value of the thermal radius is dependent on the DM properties as well and hence uncertain. Also in cold NS approximation (zero temperature), the captured DM particles need to completely lose their kinetic energy in order to be in thermal equilibrium inside the system. In an ideal though approximated case they pervade all over the NSs. Based on this discussion, we assume constant number density of the fermionic DM throughout the NS and the value to be $\sim 1000$ times smaller than the average neutron number density following \cite{Panotopoulos:2017idn}. For SNM number density $\rho_0$, the DM number density becomes roughly $\rho_{\chi} = 10^{-3}\times 10\rho_0 \sim 0.15\times 10^{-2}~\rm{fm^{-3}}$. Using this fact, we estimated a constant Fermi momentum of the DM fermions to be around $0.06~\rm{GeV}$. In absence of any proper experimental or observational evidence regarding the presence and distribution of DM in NSs, many recent works \cite{Bhat:2019tnz,%Das:2020vng,
Das:2018frc,Quddus:2019ghy}%,Das:2020ptd}
in the context of possible presence of DM in NSs, have also considered the same assumption of constant number density of DM throughout the density profile of NSs.
  
\subsection{Structural Properties of Neutron Star}
\label{Str_Prop} 
 
 With the aforesaid parameter sets for both hadronic matter and DM, we compute the EoS and structural properties of NS in presence of DM. In static condition the structural properties of NS like the central density $\rho_C$, gravitational mass $M$, baryonic mass $M_B$ and radius $R$ are obtained by solving the following Tolman-Oppenheimer-Volkoff (TOV) equations \cite{tov} numerically with the obtained EoS.
 
\begin{eqnarray}
\frac{dP}{dr}=-\frac{G}{r}\frac{\left(\varepsilon+P\right)
\left(M+4\pi r^3 P\right)}{(r-2 GM)},
\label{tov}
\end{eqnarray}

\begin{eqnarray}
\frac{dM}{dr}= 4\pi r^2 \varepsilon,
\label{tov2}
\end{eqnarray} 
 
 The baryonic mass is given by the relation
 
\begin{eqnarray} 
M_B(r)=\int_{0}^{R} 4\pi r^2 ~\rho~ m_B \left(1 - \frac{2GM}{r}\right)^{-1/2} dr
\label{barmass}
\end{eqnarray} 
 
where, $m_B$ is the mass of baryon.

 The dimensionless tidal deformability ($\Lambda$) can be obtained in terms of the mass, radius and the tidal love number ($k_2$) following \cite{Hinderer:2007mb,Hinderer:2009ca,Alvarez-Castillo:2018pve}. With the estimates of the deformation of the metric $h_{\alpha \beta}$ in Regge-Wheeler gauge,
 
\begin{eqnarray} 
h_{\alpha\beta}=diag\left[e^{-\nu(r)}H_0,e^{\lambda(r)}H_2,r^2K(r),r^2\sin^2\theta K(r)\right]
Y_{2m}(\theta,\phi)
\label{h}
\end{eqnarray}
 
the tidal Love number $k_2$ can be obtained which in turn gives the tidal deformability parameter $\lambda$ as

\begin{eqnarray} 
\lambda=\frac{2}{2} k_2 R^4
\label{lam}
\end{eqnarray}

 The dimensionless tidal deformability $\Lambda$ can then be obtained as a function of Love number, gravitational mass and radius \cite{Hinderer:2007mb,Hinderer:2009ca}.
 
\begin{eqnarray} 
 \Lambda=\frac{2}{3} k_2 (R/M)^5
 \label{Lam}
\end{eqnarray}

  The rotational properties like central density, rotational gravitational mass, baryonic mass, radius, moment of inertia $I$ and the rotational frequency $\nu$ of DM admixed NS are obtained with the help of rotating neutron star (RNS) code \cite{Stergioulas:1994ea}. 

 To calculate rotating NS properties, the RNS code \cite{Stergioulas:1994ea} is written based on the Komatsu-Eriguchi-Hachisu (KEH) method \cite{Komatsu:1989zz}. If rotation is fast, the spherical symmetry of the star is broken and it gets more and more flattened at the equator, thus obtaining a deformed (ellipsoidal) shape. However, the rotation is assumed to be continuous due to which axial symmetry is maintained. The metric for such conditions can be taken as
 
\begin{eqnarray}      
ds^2 = -e^{2 \nu}dt^2 + e^{2 \alpha} (dr^2 + r^2 d\Theta^2) +  e^{2 \beta}r^2\sin^2\Theta (d\phi - \omega dt)^2  
\end{eqnarray}         

The four velocity of rotation is given by
\begin{eqnarray}
       u^\mu = \frac{e^{-\nu}}{\sqrt{1-v^2}}(1,0,0,\Omega)  
\end{eqnarray}         

where, the proper velocity with respect to an observer with zero angular momentum is

\begin{eqnarray}
v=(\Omega-\omega)~r~\sin\Theta~e^{\beta-\nu}       
\end{eqnarray}

The limiting frequency of rotation is given by the Kepler frequency $\nu_K$, which signifies the balance between centrifugal force and gravity. A star rotating with its frequency beyond $\nu_K$ eventually becomes unstable due to the loss of mass from its equatorial region and thus it is called the mass-shedding limit. Its general relativistic expression is given as 

\begin{eqnarray}
\Omega_K = \omega + \frac{\omega'}{2\Psi'} + e^{\nu-\beta}\sqrt{\frac{\nu'}{R^2\Psi'}+\left(\frac{\omega'~e^{\beta-\nu}}{2\Psi'}\right)^2}           
\end{eqnarray}         

where, $R$ is the equatorial radius of the NS and

\begin{eqnarray}
\Psi'=\beta'+\frac{1}{R}
\end{eqnarray}       

where the prime ($'$) denotes differentiation with respect to radial coordinate $r$.

From the stress-energy tensor $T^{\mu\nu}$, the general relativistic equation for moment of inertia is given as

\begin{eqnarray}
I=\frac{2\pi}{\Omega} ~ \int dr~\int d\Theta  ~r^3\sin^2\Theta~\frac{(\varepsilon +P)v}{1-v^2}~e^{2(\alpha+\beta)}      
\end{eqnarray}                        

The rotational frequency is given by
\begin{eqnarray}   
\nu=\frac{\Omega}{2\pi}                       
\end{eqnarray}         

 These rotational quantities of NSs can be calculated using the well-known RNS code \cite{Stergioulas:1994ea}.

 In the present scenario of DM admixed NS, the rotational gravitational binding energy $BE$ is given as
 
\begin{eqnarray} 
BE=M_B + M_{DM} - M 
\label{be}
\end{eqnarray} 
  
 Since we assume the constant number density of DM fermion through constant $k_F^{\chi}$ therefore we have $M_{DM}=\frac{4}{3} \pi R^3 \rho_{\chi} m_{\chi}$. Here $M$ and $M_B$ are the rotational gravitational and baryonic mass, respectively. The rotational DM mass fraction $M_{DM}$ depends on the rotational radius of the DM admixed NS.

\section{Results and Discussions}
\label{Results}

\begin{figure}[!ht]
\centering
\includegraphics[width=0.5\textwidth]{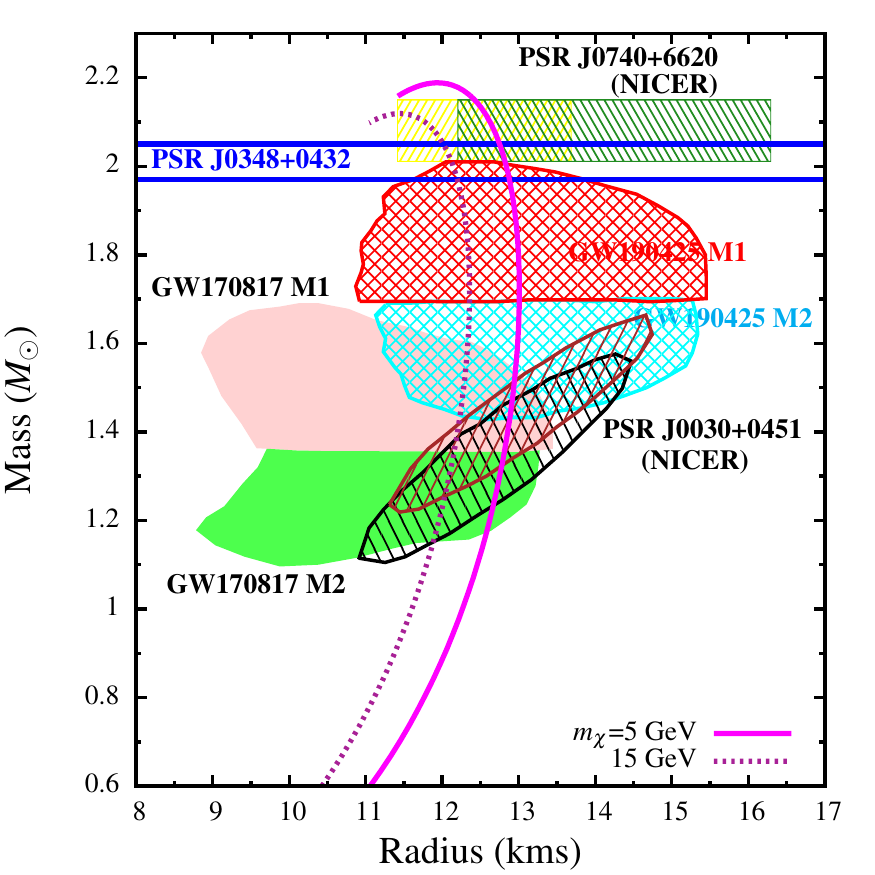}\hfill
\includegraphics[width=0.5\textwidth]{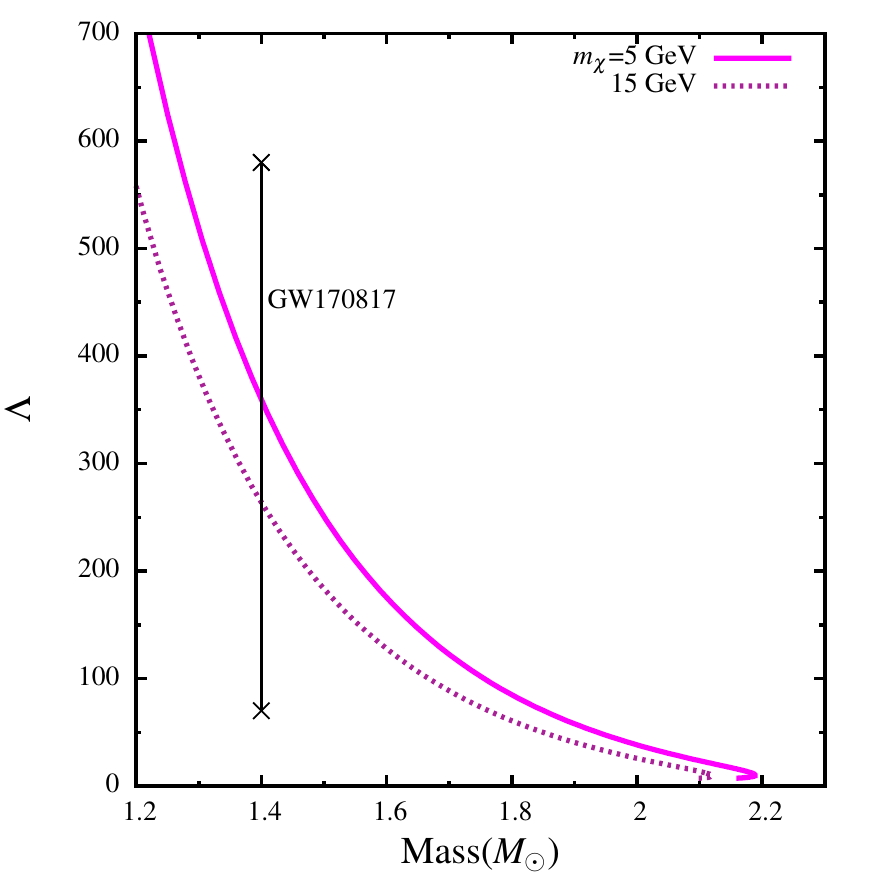}
\caption{Left: Mass-radius relationship of static dark matter admixed neutron star for different values of $m_{\chi}$. Observational limits imposed from high mass pulsars like PSR J0348+0432 ($M = 2.01 \pm 0.04 M_{\odot}$) \cite{Antoniadis:2013pzd} (blue shaded region) and PSR J0740+6620 ($M = 2.08 \pm 0.07 M_{\odot}$ \cite{Fonseca:2021wxt} and $R = 13.7^{+2.6}_{-1.5}$ km (dark green shaded region) \cite{Miller:2021qha} or $R = 12.39^{+1.30}_{-0.98}$ km (yellow shaded region) \cite{Riley:2021pdl})) are also indicated. The constraints on $M-R$ plane prescribed from GW170817 (pink (GW170817 M1) and green (GW170817 M2) shaded regions \cite{Abbott:2018exr}), GW190425 (red (GW190425 M1) and cyan (GW190425 M2) shaded regions \cite{Abbott:2020aai}) and NICER experiment for PSR J0030+0451 (black shaded region \cite{Riley:2019yda} and brown shaded region \cite{Miller:2019cac}) are also compared. Right: Variation of tidal deformability with respect to gravitational mass of static dark matter admixed neutron star for different values of $m_{\chi}$. Constraint on $\Lambda_{1.4}$ from GW170817 observations ($\Lambda_{1.4}=70-580$ \cite{TheLIGOScientific:2017qsa,Abbott:2018exr}) is also shown.}
\label{static}
\end{figure}

 For the obtained DM admixed EoS, depicted by equations \ref{e} and \ref{P}, we study the variation of gravitational mass with respect to radius in static conditions for two different values of the DM fermion mass $m_{\chi}$ and corresponding values of $m_\phi$, $y_\phi$ and $y_\xi$ as shown in table \ref{Param_DM}. The result is displayed in the left panel of figure \ref{static}. The maximum mass obtained with both the values of $m_{\chi}$ satisfy the constraints from the most massive pulsars PSR J0348+0432 \cite{Antoniadis:2013pzd} and PSR J0740+6620 \cite{Fonseca:2021wxt}. Our results are in excellent agreement with the constraints on $M-R$ plane obtained from GW170817 \cite{Abbott:2018exr} and recent NICER experiments for PSR J0030+0451 \cite{Riley:2019yda,Miller:2019cac} and PSR J0740+6620 \cite{Miller:2021qha,Riley:2021pdl}. We also studied the tidal deformation with the same EoS and the variation of dimensionless tidal deformability with respect to gravitational mass is shown in the right panel of figure \ref{static}. For both values of $m_{\chi}$, the results are in well agreement with the constraint on $\Lambda_{1.4}$ from GW170817 observations \cite{TheLIGOScientific:2017qsa,Abbott:2018exr}.
  
% We obtain the maximum mass to be $M = 2.19 M_{\odot}$ for $m_{\chi} = 5$ GeV and $M = 2.12 M_{\odot}$ for $m_{\chi} = 15$ GeV, the corresponding radius as $R = $ 11.93 km for $m_{\chi} = 5$ GeV and $R = $ 11.45 km for $m_{\chi} = 15$ GeV and the value of $R_{1.4}$ = 12.86 km for $m_{\chi} = 5$ GeV and 12.21 km for $m_{\chi} = 15$ GeV.  We found the value of $\Lambda_{1.4}$ to be 346.99 for $m_{\chi} = 5$ GeV and 256.92 for $m_{\chi} = 15$ GeV. 

 However, it can be seen that the additional effects of vector dark mediator interaction with the nucleons do not bring any significant change to the structural properties of DM admixed NSs compared to the case when only scalar DM mediator was considered in \cite{Sen:2021wev}. This is because both the scalar and vector mediator-nucleon couplings are too small and that makes the interaction strength between hadronic matter and DM too feeble to bring any noticeable change to the structural properties of DM admixed NSs. To fix the DM-mediator couplings, we assumed equal contribution of both the channels to the relic abundance, which makes this scenario somewhat similar to the case where only the scalar mediator was considered \cite{Sen:2021wev}. As seen from \cite{Sen:2021wev}, it is only the kinetic interaction of DM fermion and its mass as well as Fermi momentum that mainly contribute to the EoS and the structural properties of NSs compared to the case when only hadronic matter is considered. It is already shown in \cite{Sen:2021wev} that higher values of both $m_{\chi}$ and $k_F^{\chi}$ reduce the maximum mass, radius and tidal deformability of the DM admixed NS. 
 
 Since we do not observe any change in the DM admixed EoS on inclusion of the vector new physics mediator compared to that when only the scalar new physics mediator was considered \cite{Sen:2021wev}, it also obvious that the constraints on surface redshift from 1E 1207.4-5209 \cite{Sanwal:2002jr} and RX J0720.4-3125 \cite{Hambaryan:2017wvm} and EXO 07482-676 \cite{Cottam:2002cu} will also be satisfied for the present work.
 
 We tabulate the obtained properties of static DM admixed stars in the following table \ref{table_static}.
 
\begin{table*}[!ht]
\begin{center}
\caption{Structural properties like maximum mass, corresponding radius, $R_{1.4}$, $\Lambda_{1.4}$ and maximum surface redshift of static dark matter admixed neutron star matter for different values of $m_\chi$.}
\setlength{\tabcolsep}{15.0pt}
%{\small{
%\hline
\begin{center}
\begin{tabular}{cccccccc}
\hline
\hline
%\multicolumn{1}{c}{} &
%\multicolumn{1}{c}{$B_{as}$} &
%\multicolumn{1}{c}{$\rho_{cross}/\rho_{0}$} &
%\multicolumn{1}{c}{} &
\multicolumn{1}{c}{$m_\chi$} &
%\multicolumn{1}{c}{$\rho_{cross}/\rho_{0}$} &
\multicolumn{1}{c}{$M_{max}$} &
\multicolumn{1}{c}{$R$} &
\multicolumn{1}{c}{$R_{1.4}$} &
\multicolumn{1}{c}{$\Lambda_{1.4}$} &
\multicolumn{1}{c}{$Z_s$}\\
%\multicolumn{1}{c}{$C/m^4$} &
%\multicolumn{1}{c}{$m_{\sigma}$}\\
%\multicolumn{1}{c}{($\rm{fm^2}$)} &
%\multicolumn{1}{c}{} &
\multicolumn{1}{c}{(GeV)} &
%\multicolumn{1}{c}{} &
\multicolumn{1}{c}{($M_{\odot}$)} &
\multicolumn{1}{c}{(km)} &
\multicolumn{1}{c}{(km)} & 
\multicolumn{1}{c}{} &
\multicolumn{1}{c}{} \\
%\multicolumn{1}{c}{($\rm{fm^2}$)} &
%\multicolumn{1}{c}{(MeV)} \\
\hline
      5    &2.19    &11.93  &12.86   &346.99  &0.507 \\
      15   &2.12    &11.45  &12.21   &256.92  &0.511 \\
\hline
\hline
\end{tabular}
\end{center}
%}}
\protect\label{table_static}
\end{center}
\end{table*}

\begin{figure}[!ht]
\centering
\includegraphics[width=0.5\textwidth]{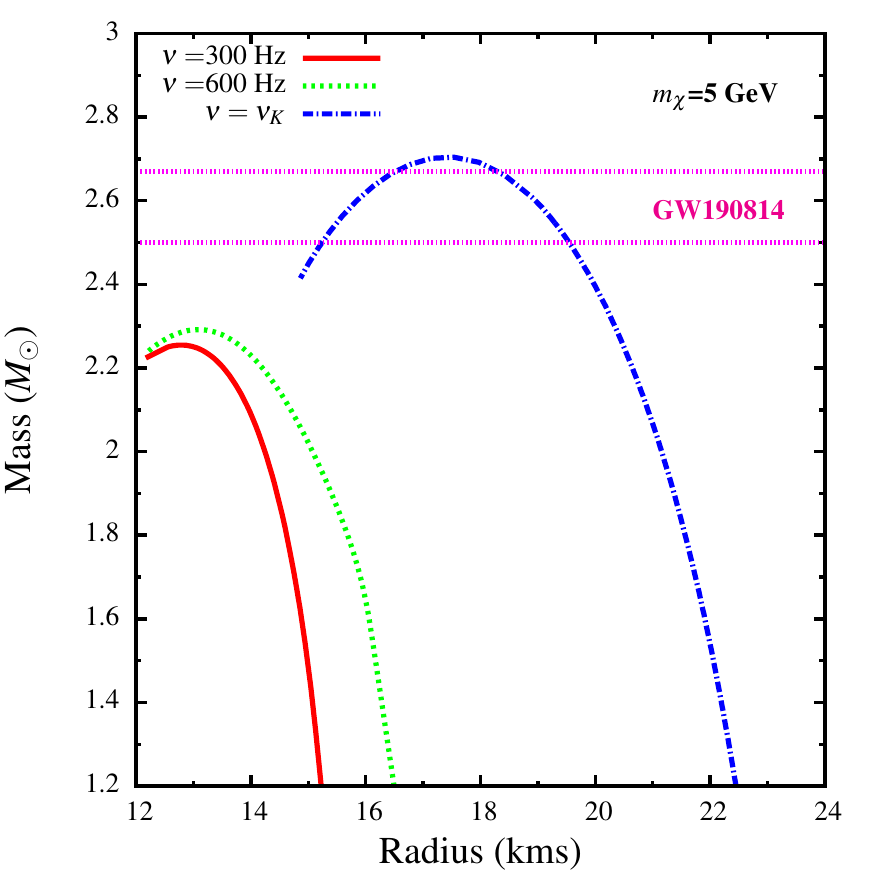}\hfill
\includegraphics[width=0.5\textwidth]{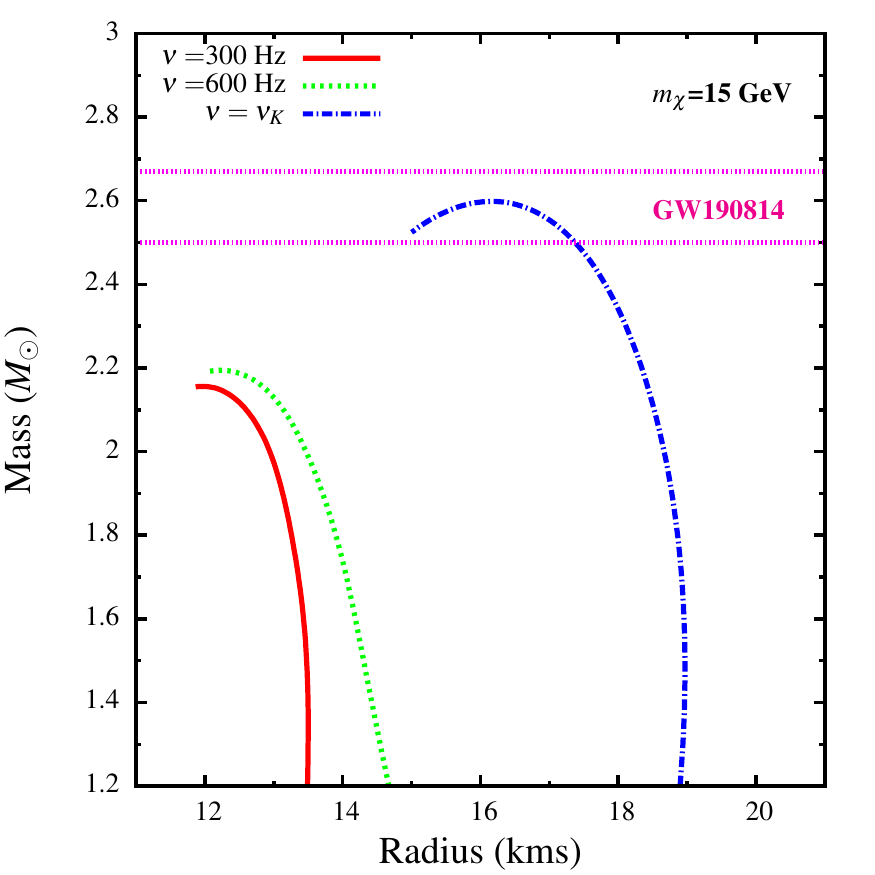}
\caption{Mass-radius relationship of dark matter admixed neutron star for different values of $m_{\chi}$ rotating with frequency $\nu=$300, 600 Hz and $\nu_K$. Mass of secondary component of GW190814 ($M = 2.59^{+0.08}_{-0.09} M_{\odot}$ \cite{Abbott:2020khf}) is also compared.}
\label{mr_rot}
\end{figure}

 We next proceed to calculate the various rotational properties of the DM admixed NSs for the same two aforesaid values of $m_{\chi}$ using the RNS code \cite{Stergioulas:1994ea}. We consider slow rotation in the present work in order to test the universality relations with our DM admixed EoS. We also calculate our results at Kepler frequency $\nu_K$ which is the maximum possible frequency for a stable rotating NS in order to account for the rapid rotation scenario. We first study the variation of gravitational mass with respect to radius for DM admixed NS rotating at frequency $\nu=$ 300 and 600 Hz (angular velocity $\Omega=$ 0.19 and 0.38 $\times$ 10$^{4}$ s$^{-1}$, respectively) and at $\nu_K$. In figure \ref{mr_rot} we present the results for the same two values of $m_{\chi}$ as chosen for the static case. As expected, irrespective of the value of $m_{\chi}$, the gravitational mass and radius increase for rotating NSs compared to that of the static ones. This is because the the centrifugal force has a very important role to play in the former case. This centrifugal force becomes stronger with the increase of $\nu$ and therefore we obtain more massive NSs with larger radius for a faster rotating NS.

  At Kepler frequency our results of maximum rotational mass matches with the mass of the secondary component of GW190814 \cite{Abbott:2020khf} for both the chosen values of $m_{\chi}$ indicating that this object may be a fast rotating DM admixed NS.
 
\begin{figure}[!ht]
\centering
\includegraphics[width=0.5\textwidth]{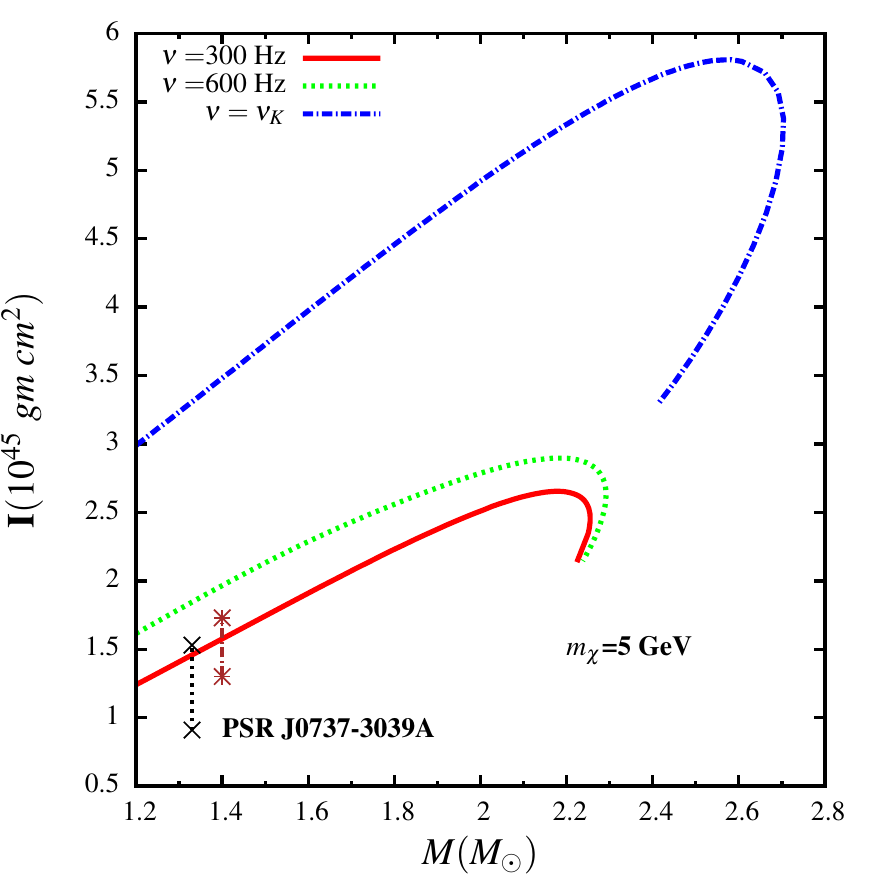}\hfill
\includegraphics[width=0.5\textwidth]{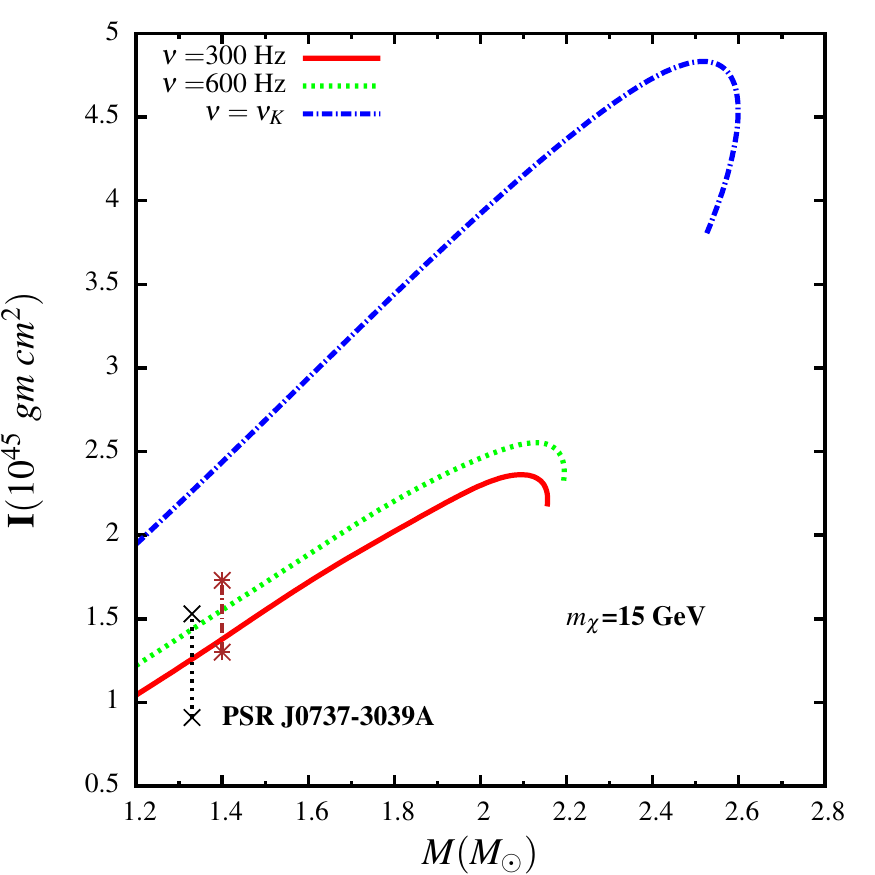}
\caption{Variation of moment of inertia with respect to gravitational mass of dark matter admixed neutron star for different values of $m_{\chi}$ rotating with frequency $\nu=$300, 600 Hz and $\nu_K$. Constraint from PSR J0737-3039A ($I_{1.338}=1.15^{+0.38}_{-0.24}$ $\times 10^{45}$ g cm$^{2}$ \cite{Landry:2018jyg}) is shown by the black dotted line with crossmarks and that from \cite{Jiang:2019rcw} ($I_{1.4}=1.43^{+0.30}_{-0.13}$ $\times 10^{45}$ g cm$^{2}$) is shown with brown dashed line with asterisks.}
\label{mI_rot}
\end{figure}

 The moment of inertia $I$ is calculated next for the same conditions and its variation with gravitational mass shown in figure \ref{mI_rot}. The moment of inertia also increases with rotational speed for a given DM admixed EoS. This is because moment of inertia is larger for massive stars as they can sustain faster rotation. Moreover, for any particular rotational frequency, the gravitational mass is more for lower values of $m_{\chi}$ (figure \ref{mr_rot}). Therefore we obtain larger moment of inertia for $m_{\chi}=$ 5 GeV compared to that for $m_{\chi}=$ 15 GeV.

  For $m_{\chi}=$ 5 GeV the constraints on $I_{1.338}$ from PSR J0737-3039A \cite{Landry:2018jyg} and $I_{1.4}$ \cite{Jiang:2019rcw} are satisfied with rotational frequency $\nu=$ 300 Hz while for $m_{\chi}=$ 15 GeV the same constraints are satisfied with both $\nu=$300 and 600 Hz. For slow rotation ($\nu=$300 and 600 Hz), the moment of inertia for both the values of $m_{\chi}$ are consistent with the range ($1.0-3.6 \times 10^{45}$ gm cm$^2$) specified by \cite{Bhattacharyya:2017tos} for slow rotation.
 
\begin{figure}[!ht]
\centering
\includegraphics[width=0.5\textwidth]{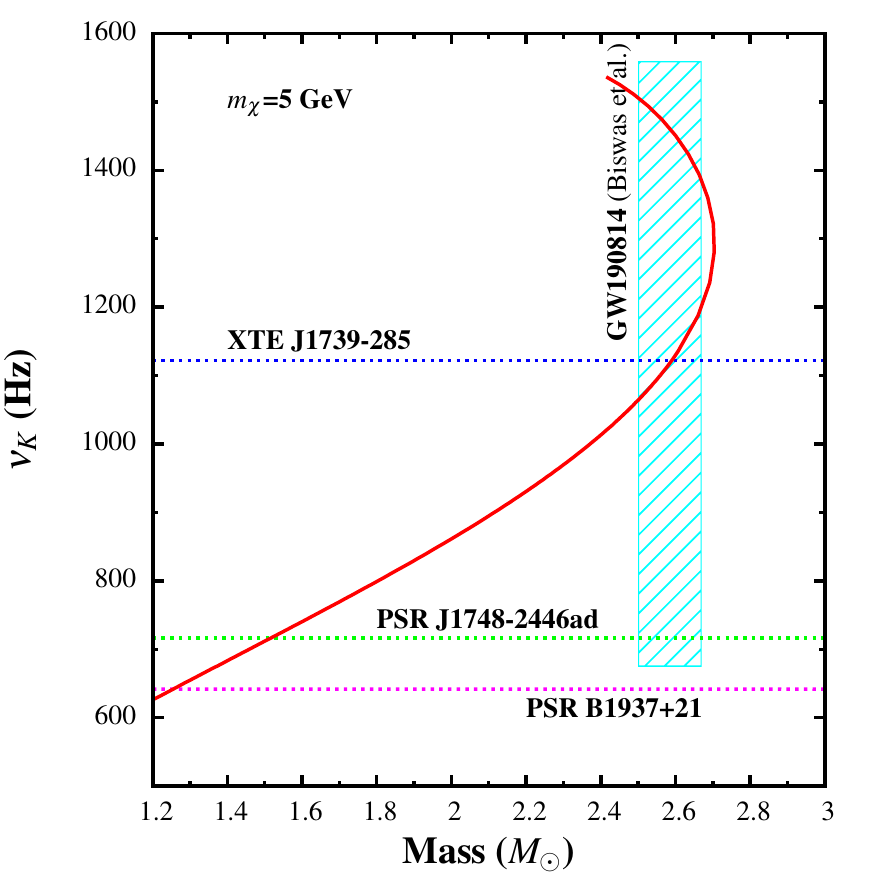}\hfill
\includegraphics[width=0.5\textwidth]{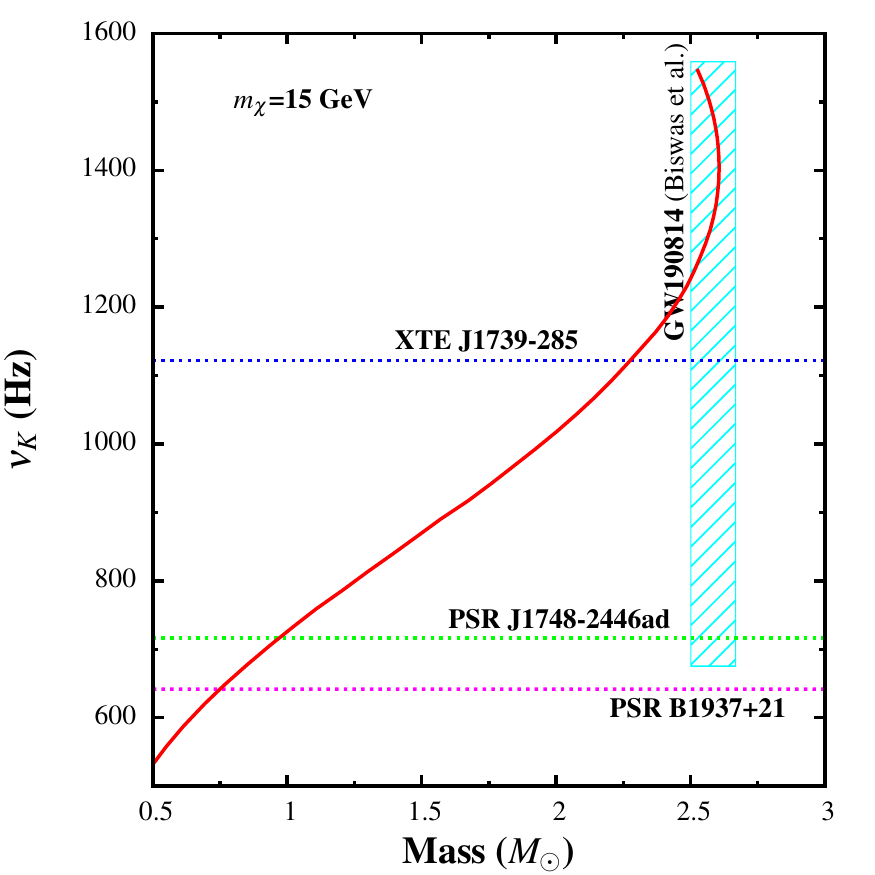}
\caption{Variation of rotational frequency with respect to gravitational mass of dark matter admixed neutron star for different values of $m_{\chi}$ rotating at Kepler frequency velocity. The frequencies from fast rotating pulsars such as PSR B1937+21 ($\nu = 633$ Hz) \cite{Backer} and PSR J1748-2446ad ($\nu = 716$ Hz) \cite{Hessels:2006ze} and XTE J1739-285 ($\nu=1122$ Hz) \cite{Kaaret:2006gr} are also indicated. Result of \cite{Biswas:2020xna} ($\nu = 1170^{+389}_{-495}$ - cyan shaded region) for the secondary component of GW190814 is also compared.}
\label{mn_rot}
\end{figure}

 In figure \ref{mn_rot} we show the rotational frequency profile obtained with variation of central density at Keplerian velocity. As expected the rotational frequency increases with mass irrespective of the value of $m_{\chi}$.

  The constraints on $\nu$ from PSR B1937+21 \cite{Backer} and PSR J1748-2446ad \cite{Hessels:2006ze} are satisfied for both the values of $m_{\chi}$. It is also known that the mass of the pulsar PSR J1748-244ad is $< 2 M_{\odot}$ rotating at 716 Hz \cite{Hessels:2006ze}. Our results are in well agreement with the same to a good extent for both the values of $m_{\chi}$. The constraint from XTE J1739-285 \cite{Kaaret:2006gr} is not yet confirmed. However, we have satisfied it for both the values of $m_{\chi}$. Moreover, our results are also consistent with the very recent findings of \cite{Biswas:2020xna} that gives a probable estimate of the rotational frequency of the secondary component of GW190814.
 
 Thus from figures \ref{mr_rot} and \ref{mn_rot} it can be said that the secondary component of GW190814 may be a fast rotating DM admixed NS.

\begin{figure}[!ht]
\centering
\includegraphics[width=0.5\textwidth]{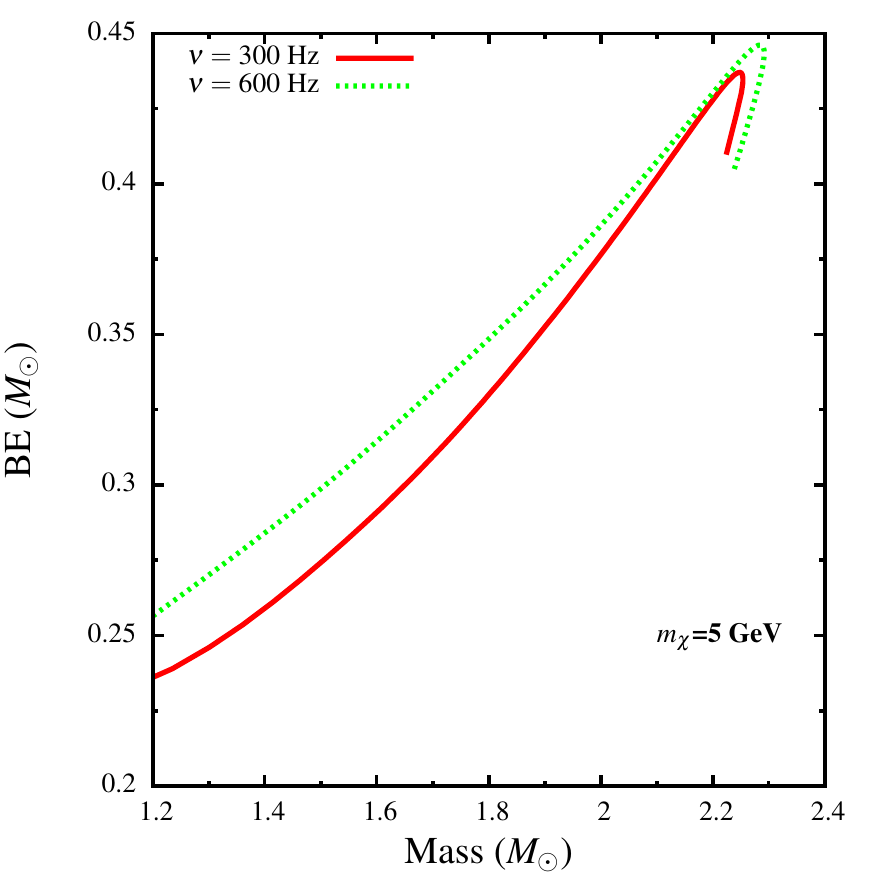}\hfill
\includegraphics[width=0.5\textwidth]{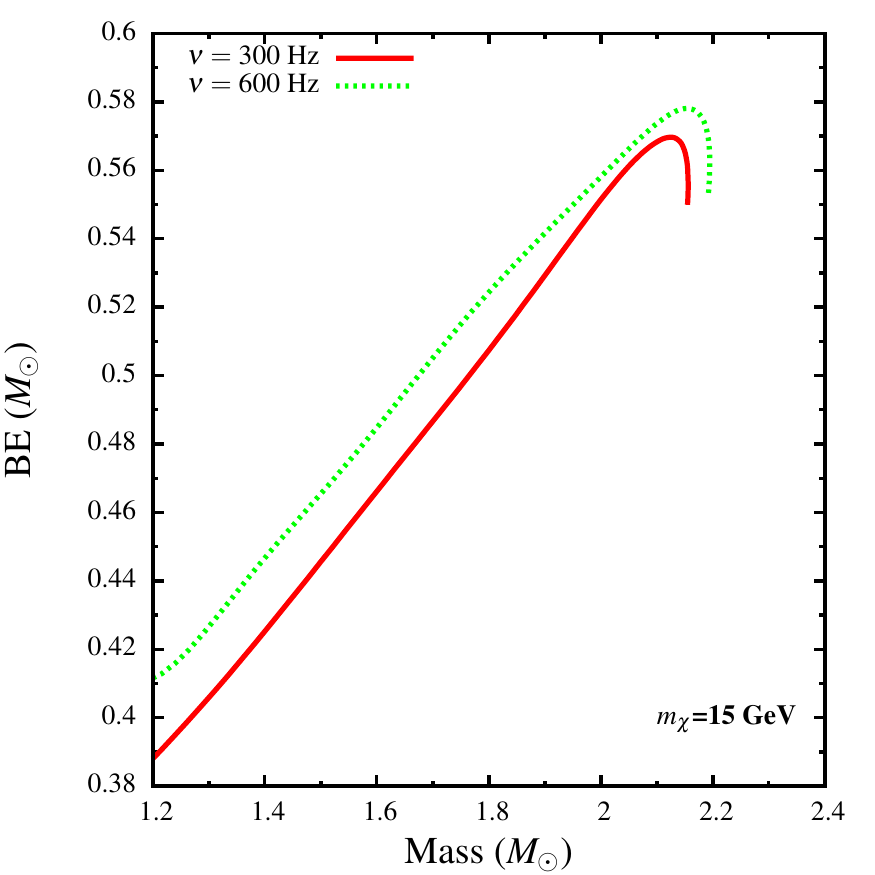}
\caption{Variation of gravitational binding energy with respect to gravitational mass of dark matter admixed neutron star for different values of $m_{\chi}$ rotating slowly with frequency $\nu=$300 and 600 Hz.}
\label{be_rot}
\end{figure}

 Considering slow rotation ($\nu=$300 and 600 Hz), we now calculate the gravitational binging energy \cite{Bhattacharyya:2017tos,Bhat:2018ayy} of the DM admixed NSs for the two chosen values of $m_{\chi}$. The gravitational binging energy $BE$ is the difference between the baryonic mass $M_B$ and gravitational mass $M$ of a star. However, in the present scenario that considers the presence of DM in NS, the $BE$ is calculated following equation \ref{be} and in figure \ref{be_rot} we show the variation of $BE$ of the slowly rotating DM admixed NS.

  For $m_{\chi}=$ 5 GeV our estimates of $BE$ for both the chosen values $\nu$ are well consistent with the range $(0.1 - 0.5)M_{\odot}$ obtained by \cite{Bhattacharyya:2017tos} for various theoretical models. For $m_{\chi}=$ 15 GeV, the value of $BE$ overshoots a bit. Since we have considered constant number density of DM ($\rho_{\chi}$), more massive DM fermions produce more mass density of DM as well as more mass fraction ($M_{DM}$) of DM inside NS. Naturally gravitational binding energy can be expected to be higher in the latter case. Also over the decades of development in the direct detection experiments ruled out a large parameter space for $m_{\chi} \ge 10~\rm{GeV}$. Very recent findings of excess events around $2~\rm{keV}$ recoil energy range, suggest that it is very unlikely to find DM particle candidates of mass $m_{\chi} \ge 10~\rm{GeV}$~\cite{Aprile:2018dbl}.
 
\begin{figure}[!ht]
\centering
\includegraphics[width=0.5\textwidth]{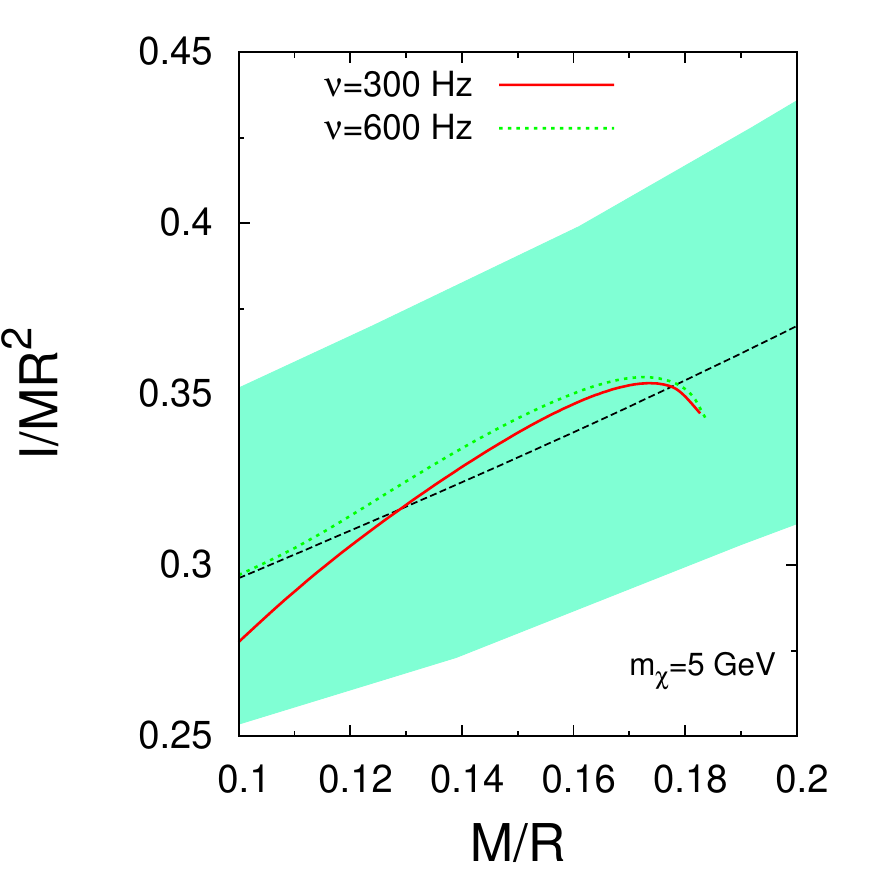}\hfill
\includegraphics[width=0.5\textwidth]{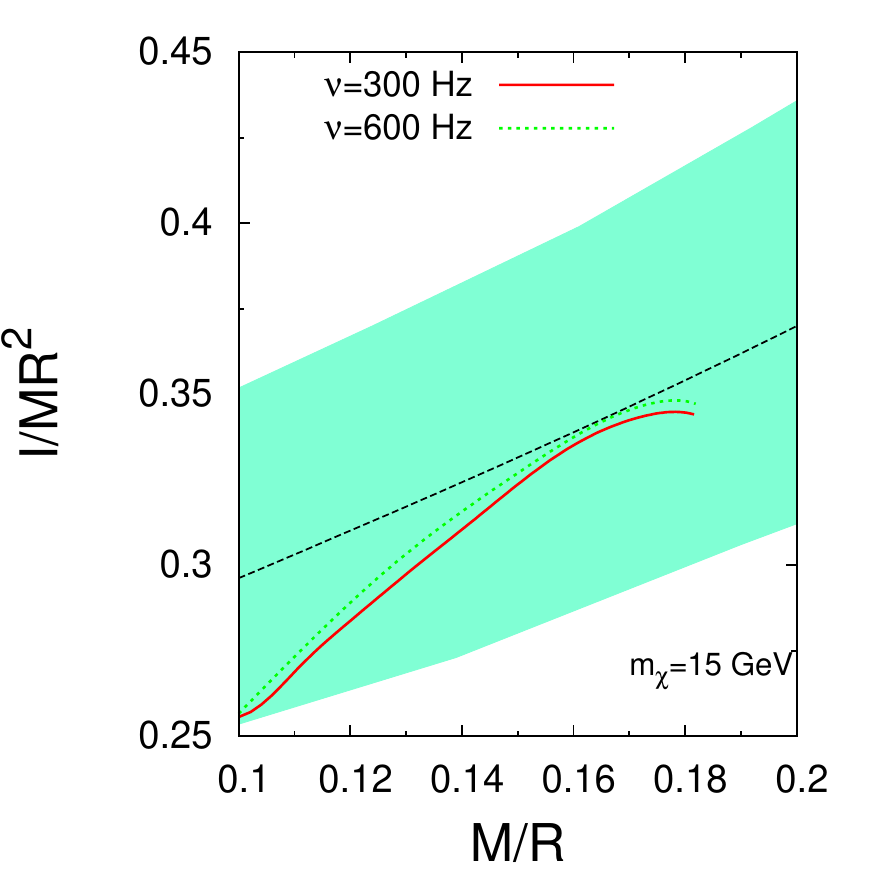}
\caption{Normalized moment of inertia ($I/MR^2$) versus compactness factor ($M/R$) of dark matter admixed neutron star for different values of $m_{\chi}$ rotating slowly with frequency $\nu=$300 and 600 Hz. The fitted value of normalized $I$ from various theoretical models for slow rotation (black dashed line) \cite{Lattimer:2004nj} is shown along with the uncertainty region (shaded region) \cite{Breu:2016ufb}.}
\label{IMR2_rot}
\end{figure}

\begin{figure}[!ht]
\centering
\includegraphics[width=0.5\textwidth]{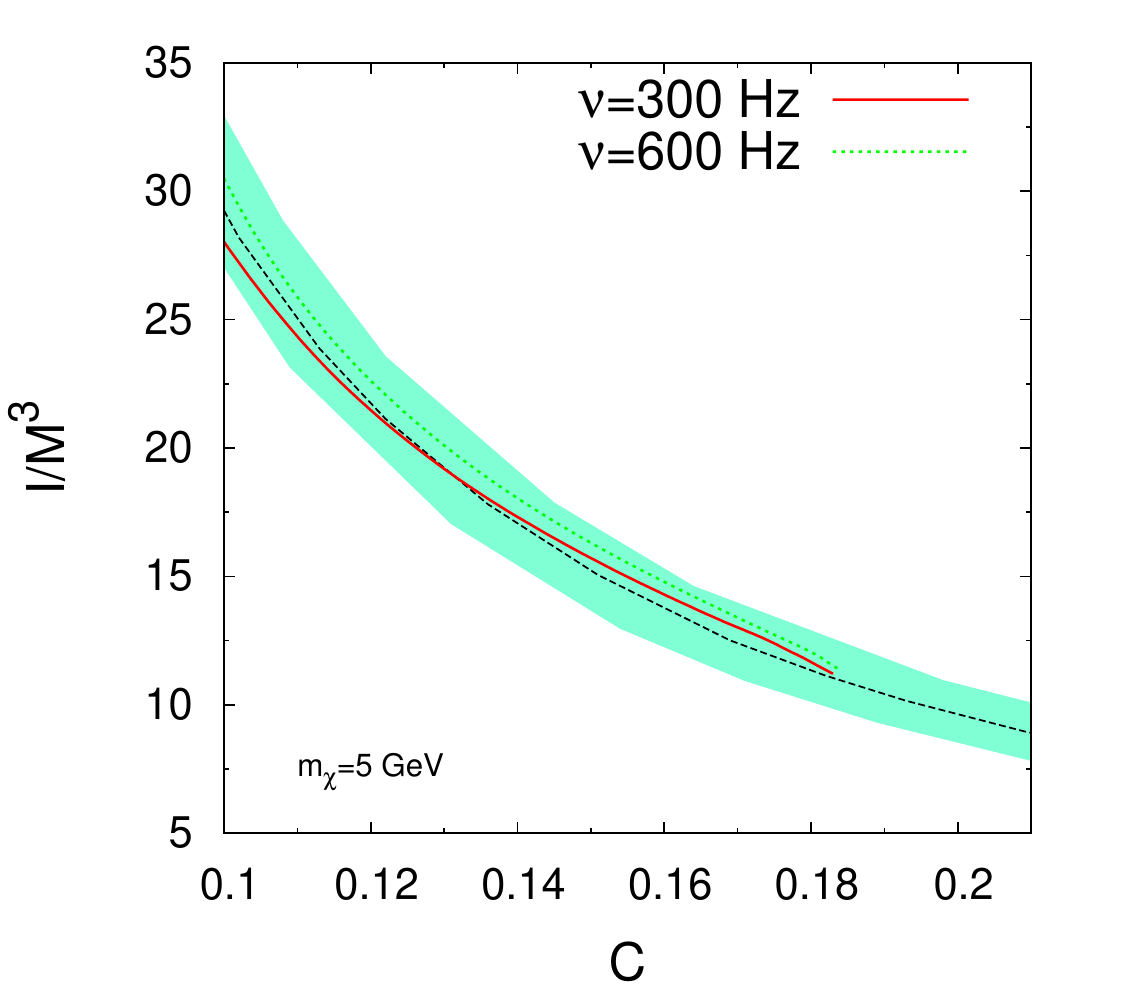}\hfill
\includegraphics[width=0.5\textwidth]{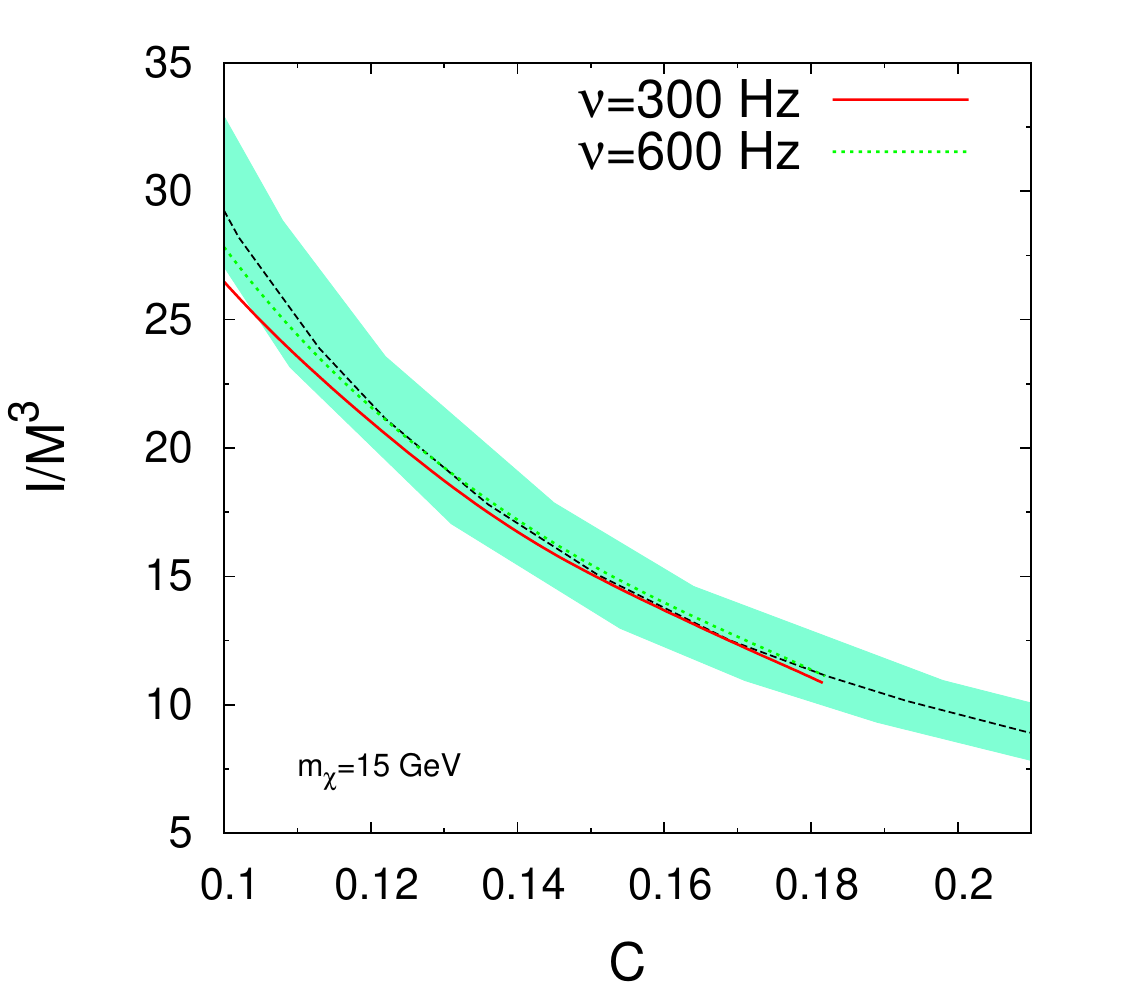}
\caption{Normalized moment of inertia ($I/M^3$) versus compactness factor ($C=M/R$) of dark matter admixed neutron star for different values of $m_{\chi}$ rotating slowly with frequency $\nu=$300 and 600 Hz. The fitted value of normalized $I$ from various theoretical models for slow rotation (black dashed line) \cite{Lattimer:2004nj} is shown along with the uncertainty region (shaded region) \cite{Breu:2016ufb}.}
\label{IM3_rot}
\end{figure}

 We now employ our DM admixed EoS for the same two chosen values of $m_{\chi}$ to calculate the normalized moment of inertia ($I/MR^2$ and $I/M^3$) of slowly rotating ($\nu=$300 and 600 Hz) NSs. In order to test the universality of the obtained DM admixed EoS in the present work, we show the variation of $I/MR^2$ and $I/M^3$ with respect to compactness parameter $C(=M/R)$ \cite{Breu:2016ufb,Lattimer:2004nj,Yagi:2013awa,Sen:2018yyq,Sen:2021bms} in figures \ref{IMR2_rot} and \ref{IM3_rot}, respectively.

 It is seen from both figures \ref{IMR2_rot} and \ref{IM3_rot} that in terms of the normalized moment of inertia, the universality of our DM admixed EoS for both the values of $m_{\chi}$ holds quite well with respect to the theoretical constraints set by \cite{Breu:2016ufb,Lattimer:2004nj} for slow rotation from different models. It is also found that our estimates of $I/MR^2$ for both the chosen values of $m_{\chi}$ and $\nu$ are consistent with the range ($0.38 \pm 0.05$) obtained by \cite{Bhattacharyya:2017tos} for different models.

 The obtained rotational properties are tabulated below in table \ref{table_rot}. 
 
\begin{table*}[!ht]
\begin{center}
\caption{Structural properties like maximum mass, corresponding radius and maximum moment of inertia of dark matter admixed neutron star matter for different values of $m_\chi$ rotating at different rotational frequencies $\nu$.}
\setlength{\tabcolsep}{15.0pt}
%{\small{
%\hline
\begin{center}
\begin{tabular}{cccccccc}
\hline
\hline
%\multicolumn{1}{c}{} &
%\multicolumn{1}{c}{$B_{as}$} &
%\multicolumn{1}{c}{$\rho_{cross}/\rho_{0}$} &
%\multicolumn{1}{c}{} &
\multicolumn{1}{c}{$m_\chi$} &
\multicolumn{1}{c}{$\nu$} &
\multicolumn{1}{c}{$M_{max}$} &
\multicolumn{1}{c}{$R$} &
\multicolumn{1}{c}{$I_{max}$} \\
%\multicolumn{1}{c}{$C/m^4$} &
%\multicolumn{1}{c}{$m_{\sigma}$}\\
%\multicolumn{1}{c}{($\rm{fm^2}$)} &
%\multicolumn{1}{c}{} &
\multicolumn{1}{c}{(GeV)} &
\multicolumn{1}{c}{(Hz)} &
\multicolumn{1}{c}{($M_{\odot}$)} &
\multicolumn{1}{c}{(km)} &
\multicolumn{1}{c}{($\times 10^{45}$ gm cm$^2$)}  \\
%\multicolumn{1}{c}{($\rm{fm^2}$)} &
%\multicolumn{1}{c}{(MeV)} \\
\hline
      5    &300        &2.25   &12.78   &2.65 \\
           &600        &2.30   &13.11   &2.90 \\
           &$\nu_K$    &2.70   &17.14   &5.81 \\    
\hline         
      15   &300        &2.16   &12.03   &2.35 \\
           &600        &2.20   &12.29   &2.55 \\
           &$\nu_K$    &2.61   &16.27   &4.83 \\    
\hline
\hline
\end{tabular}
\end{center}
%}}
\protect\label{table_rot}
\end{center}
\end{table*}

%\newpage
\section{Summary and Conclusion}
\label{Conclusion}
 
 The present work is an extension of our previous work \cite{Sen:2021wev} where we introduced a light new physics scalar mediator for the feeble interaction between hadronic matter and DM in static NSs. In the present work in addition to this scalar mediator, we introduce a vector new physics mediator to study its effects on the DM admixed EoS and consequently the structural properties of the DM admixed NS. It is noteworthy that the masses of DM fermion, the mediators and the couplings are adopted from the self-interaction constraint from Bullet cluster and from present day relic abundance. However, in the present work we find that the vector mediator do not bring any considerable effect the DM admixed EoS that was obtained considering only the scalar new physics mediator. Consequently, we do not observe any perceptible change in the static properties of the DM admixed NSs due to considering the vector new physics mediator for the chosen values of DM fermion mass and the couplings. The structural properties obtained in static case like the maximum gravitational mass and radius and tidal deformability of a 1.4 $M_{\odot}$ NS are found to be decreasing with increasing value of $m_{\chi}$ for a particular constant value of Fermi momentum of DM fermion. The obtained structural properties are in good agreement with the various constraints on them from different perspectives like the detection of massive pulsars like PSR J0348+0432 and PSR J0740+6620, the gravitational wave GW170817 from BNSM and the recently obtained results of NICER experiments for PSR J0030+0451 and PSR J0740+6620. The observational bounds on maximum redshift obtained from 1E 1207.4-5209 and RX J0720.4-3125 is also satisfied by the DM admixed EoS of the present work.
 
  We also extended the present work to calculate the properties of the DM admixed NS in rotating conditions. We studied and obtained the results considering both slow and rapid rotation. The rotational mass and radius are found to be more than that in static case. Also, they both along with the moment of inertia attain higher values with increasing rotational frequency. At Keplerian velocity, with our DM admixed EoS we not only satisfy the constraint on maximum mass from the secondary component of GW190814 but also its predicted range of rotational frequency by \cite{Biswas:2020xna} considering it to be a rapidly rotating pulsar. Thus it may be a fast rotating massive DM admixed NS. Our estimates of rotational frequency at Keplerian velocity also satisfy the constraints from PSR B1937+21, PSR J1748-2446ad and XTE J1739-285. 
  
  In slowly rotational conditions the moment of inertia obtained also agree with the  
constraints from PSR J0737-3039A and that on $I_{1.4}$. We also found that for our DM admixed EoS, the universality relations hold quite well in terms of normalized moment of inertia.

\end{document}